\pdfoutput=1
\documentclass{iau}
\usepackage{graphicx}
\newcommand{\hii}{H~\textsc{ii}}
\newcommand{\msun}{$M_\odot$}

\title[Dramas of HII~regions] 
{Drama of HII~regions: Clustered and Triggered Star Formation}

\author[Li et al.]   
{Jin-Zeng Li$^1$, Jinghua Yuan$^1$, Hong-Li Liu$^1$,
Yuefang Wu$^2$, \and Ya-Fang Huang$^1$}

\affiliation{$^1$National Astronomical Observatories, Chinese Academy of Sciences, \\ 20A Datun Road, Chaoyang District, Beijing 100012, China \\ email: {\tt ljz@nao.cas.cn} \\[\affilskip]
$^2$Department of Astronomy, Peking University, 100871 Beijing, China }

\pubyear{2015}
\volume{316}  
\jname{Formation, evolution, and survival of massive star clusters}
\editors{C. Charbonnel \& A. Nota, eds.}
\begin{document}

\maketitle

\begin{abstract}
In order to understand the star formation process under the influence of \hii~regions, we have carried out extensive investigations to well selected star-forming regions which all have been profoundly affected by existing massive O type stars. On the basis of multi-wavelength data from mid-infrared to millimeter collected using $Spitzer$, $Herschel$, and ground based radio telescopes, the physical status of interstellar medium and star formation in these regions have been revealed. In a relatively large infrared dust bubble, active star formation is undergoing and the shell is still expanding. Signs of compressed gas and triggered star formation have been tentatively detected in a relatively small bubble. The dense cores in the Rosette Molecular Complex detected at 1.1 mm using SMA have been speculated to have a likely triggered origin according to their spatial distribution. Although some observational results have been obtained, more efforts are necessary to reach trustworthy conclusions. 
\keywords{ISM: bubbles -- \hii~regions -- stars: formation -- stars: massive}
\end{abstract}

\noindent \textbf{1. N6 -- an expanding large bubble with active star formation (\cite[Yuan et al. 2014]{yuan14})}

\begin{figure}[b]
\begin{center}
 \includegraphics[width=0.7\textwidth]{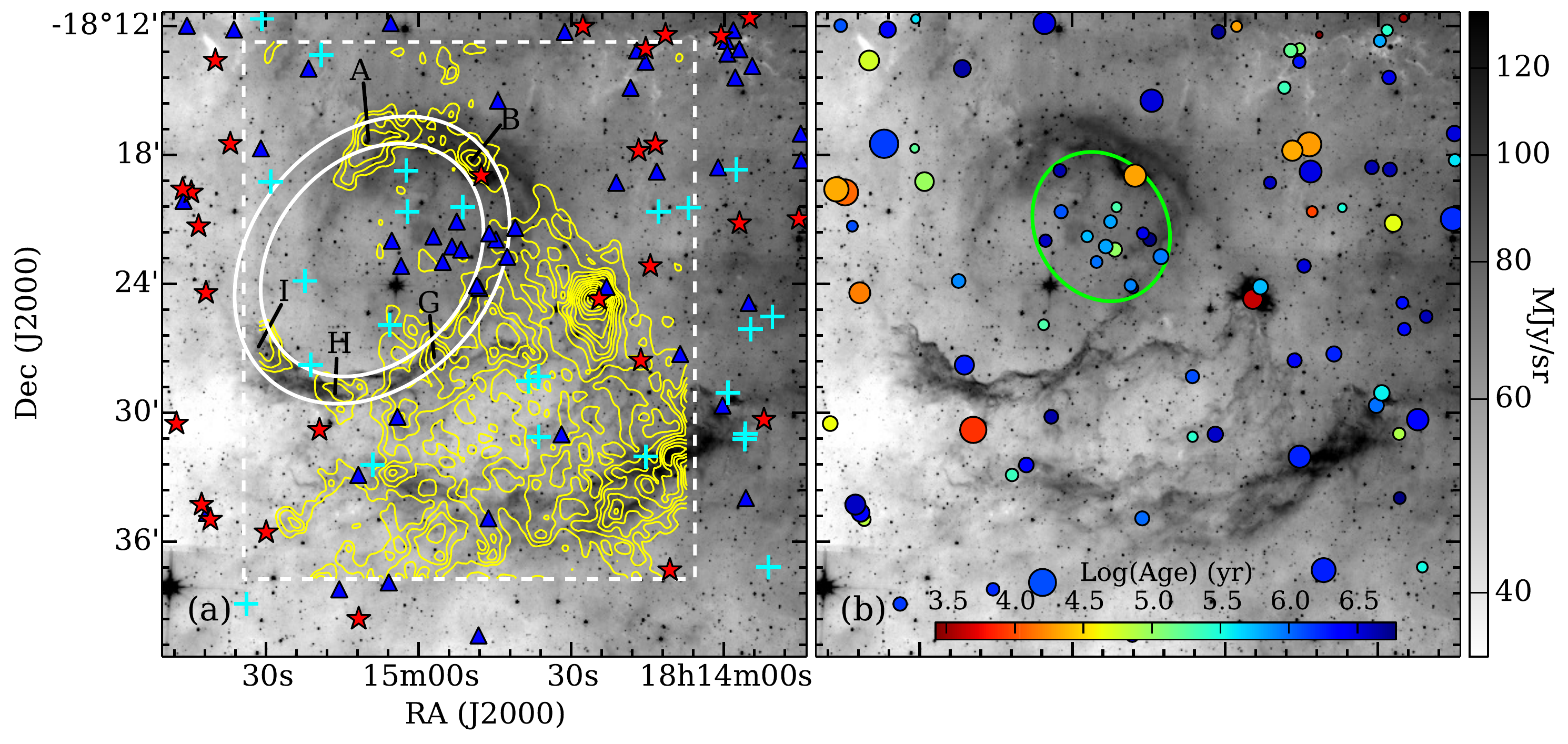}
 \caption{Appearance of the bubble N6 revealed in the infrared (background) and molecular line (contours) emission. Different symbols represents identified YSOs whose ages are color coded and masses are indicated by symbol sizes in the right panel.
}
   \label{n6}
\end{center}
\end{figure}
Although with a diameter of 11 pc, the infrared bubble N6 is still in expansion and actively giving birth to new stars, which have been revealed by the following observational features. A shell structure can be clearly traced by both infrared continuum and molecular line emission (Figure \ref{n6}). A velocity gradient along the NW-SE orientation for the ring structure has been evidently detected. Among the ten detected molecular clumps, five (A, B, G, H, and I) are located on the border of the ring structure, and four of them show elongations along the shell, consistent with a collect and collapse scenario. Based on the number of YSOs with masses of $3-8$ \msun, the lower limits of number and surface density of YSOs inside the bubble are estimated to be 770 and 21 pc$^{-2}$, respectively, indicating active star formation.

\noindent\textbf{2. G24.134 -- compressed gas and triggered star formation (\cite[Liu et al. 2015]{liu15})}

Based on multi-wavelength investigations, the Galactic bubble G24.136 is found to be compressing its ambient ISM and have the potential to trigger the formation of stars. Dense molecular gas has been detected along the shell structure (Figure 2). The probability distribution function of the column densities can be well fitted using two log-normal functions (Figure 2). Such a  feature may be arising from the turbulent cloud and the compressed layers. Six molecular clumps have been found to be along the bubble shell, resembling the scenario of collect and collapse (Figure 2).

\begin{figure}[thb]
\begin{center}
 \includegraphics[width=0.9\textwidth]{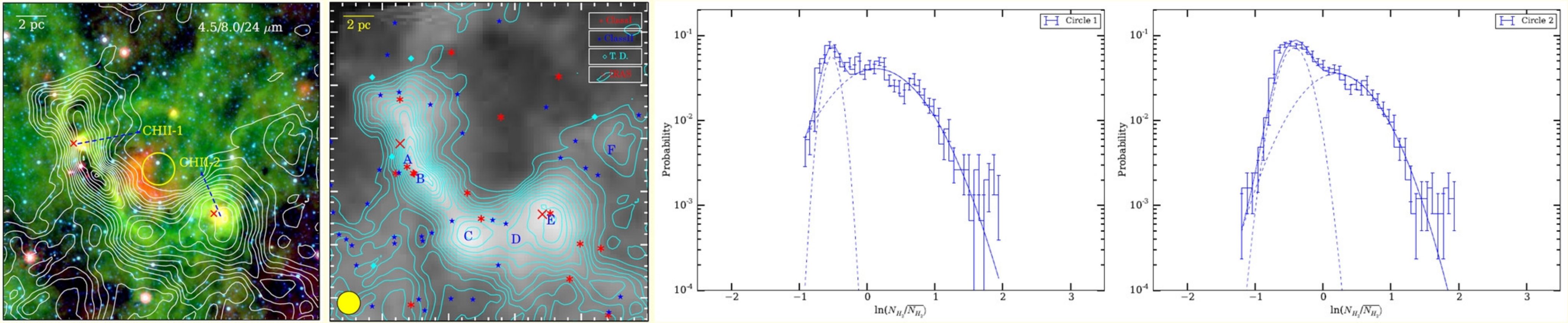}
 \caption{$Left:$ Appearance of the bubble G24.136 revealed in the infrared (background) and molecular line (contours) emission.  $Right$: Column density PDFs of the selected regions which can be well fitted with the combination of two log-normal distributions.
}
   \label{g24}
\end{center}
\end{figure}

\noindent \textbf{3. AFGL 961 -- shaping regional core mass function (\cite[Li et al. 2013]{li13})}

In the ridge of  the Rosette Molecular Complex, the SMA observations have revealed more than 40 pre- and proto-stellar cores which are preferentially located near the massive molecular outflows originating from the central massive binary (Figure 3).
The differential core mass function is computed, which yields a somewhat shallower slope of 1.8 as compared to that of the stellar IMF (Figure 3). Dynamical processes from the newly formed massive YSOs could have changed the initial conditions of core or star formation and enhanced the formation of more massive cores near the center of the cluster.

\begin{figure}[thb]
\begin{center}
 \includegraphics[width=0.7\textwidth]{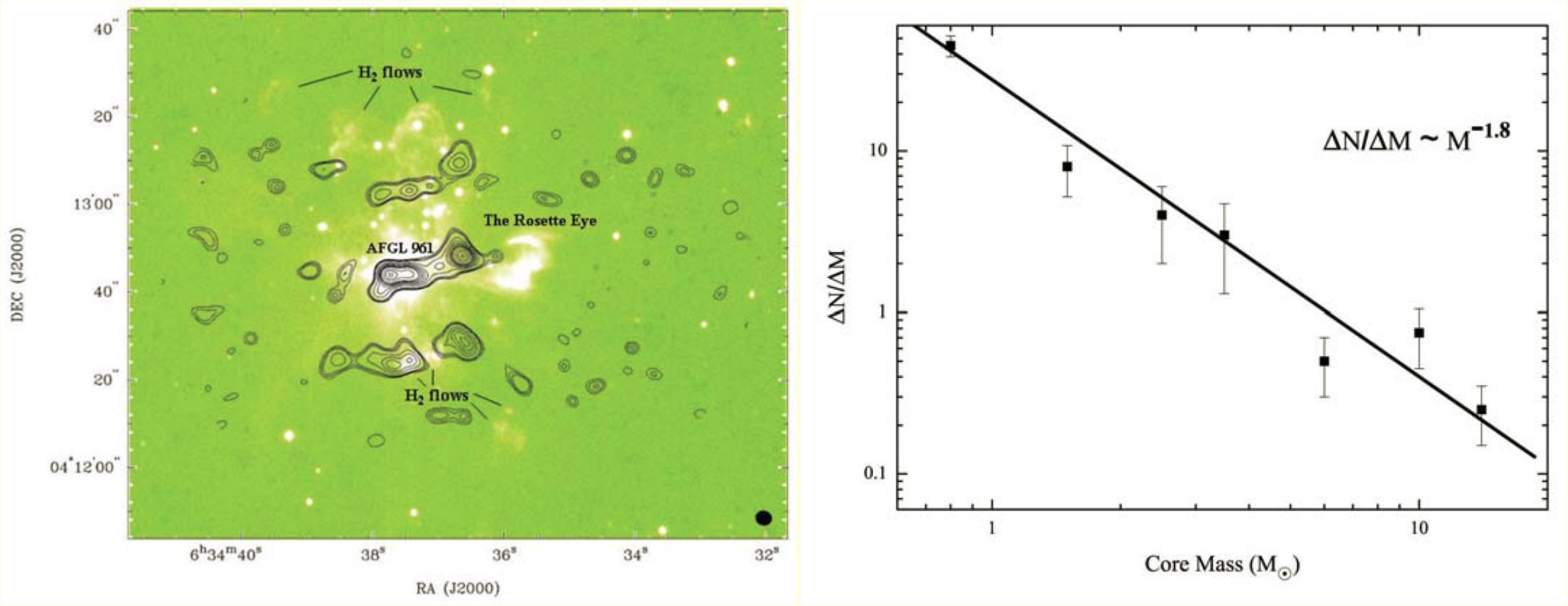}
 \caption{$Left$: Submillimeter  contours  overlaid onto NTT H$_2~1-0$~S(1) image of the AFGL 961 region. $Right$: Differential core mass function
}
\label{rosette}
\end{center}
\end{figure}

\end{document}